# Topic Detection from Conversational Dialogue Corpus with Parallel Latent Dirichlet Allocation Model and Elbow Method

Haider Khalid[1] and Vincent Wade[2]

[1]School of Computer Science and Statistics, Trinity College Dublin, University of Dublin, Dublin, Ireland
[2]ADAPT Centre, Trinity College Dublin, University of Dublin, Dublin, Ireland

## Abstract

*A conversational system needs to know how to switch between topics to continue the conversation for a more extended period. For this topic detection from dialogue corpus has become an important task for a conversation and accurate prediction of conversation topics is important for creating coherent and engaging dialogue systems. In this paper, we proposed a topic detection approach with Parallel Latent Dirichlet Allocation (PLDA) Model by clustering a vocabulary of known similar words based on TF-IDF scores and Bag of Words (BOW) technique. In the experiment, we use K-mean clustering with Elbow Method for interpretation and validation of consistency within-cluster analysis to select the optimal number of clusters. We evaluate our approach by comparing it with traditional LDA and clustering technique. The experimental results show that combining PLDA with Elbow method selects the optimal number of clusters and refine the topics for the conversation.*

## 1. Introduction

Almost fifty years ago, ELIZA [16] was created as the first conversational software and considered as an intelligent chat-bot. It was intended to emulate psycho-therapist. Today, conversational systems emerging in many domains, ranging from ticket reservation to educational context. In recent years, many conversational systems have been introduced, such as Google Assistant, Amazon Alexa, Apple Siri, Microsoft Cortana. Conversational agents often restrict for task-oriented systems [8] to achieve a particular task but human conversational dialogue is however, too complex to be handled by a simple intent-based system [5]. The system requires more human conversational dialogues become more complicated to understand the context of the conversation and decide what to say next depends on anything more than the current user input. There are mainly four categories for intelligent conversational systems are task-oriented, answering the questions, social conversational system and purposeful conversational systems. Natural Language Understanding (NLU) is the central aspect of intelligent systems. In NLU tasks, we can extract meaning from words, sentences, paragraph and a document. There is a series of hierarchy to extract the context of a conversation. At the dialogue utterance level, one of the most useful ways to understand the text is by analysing its topics and the context of the conversation.

In this paper, we combine the Parallel Latent Dirichlet Allocation (PLDA) Model with K-mean clustering technique. Clustering vocabulary of known similar words based on TF-IDF scores and Bag of Words (BOW) approach. In this approach, each dialogue converted as a document in the pre-processing data phase. Using the classical bag of words approach with TF-IDF weighting scheme dialogues are represented. The similarity measure is used for clustering the combination of document-to-document and document-to- cluster. Also, we use the elbow method for interpretation and validation of consistency within-cluster analysis to select the optimal number of clusters. In order to study the performance of semantic similarity between similar words, noise is removed from data pre-processing phase, we use precision, recall and F-measures for the evaluation and compared our results with traditional LDA and clustering technique.

In section 2, we describe topic detection challenges in a dialogue system. Also, how topic detection is different from the dialogue system as compared with topic detection from tweets, blogs and textual documents. Section 3, mentions the state of art-related work on topic detection and existing techniques for topic detection from textual data. Section 4, explains the proposed approach and methods used for the experiment. Section 5, provides the experimental results and the evaluation metrics together with the comparison between existing approaches and the proposed approach.

## 2. TOPIC DETECTION CHALLENGES IN DIALOGUE SYSTEMS

The significant difference between topic detection in textual documents/tweets and dialogue corpus is textual documents or tweets are stable data and not changing in context with different times. But conversational dialogues change the context of the conversation over time. For example, yesterday it was raining, and we collect the conversational data on weather information and train our system on this weather data. Today weather condition is different and when a user starts a conversation with the system and system detects the topic about the weather. System responses on the weather condition will be different according to today's weather conditions. In results, the system performs correctly according to the training mechanism, but the system will lose user engagement during the conversation.

The conversational dialogues are short text with irregular writing styles, abbreviations and synonyms. An incremental clustering technique helps to find similarities and topic detection in a temporal context [10]. Contextual topic modelling is also the main challenge in a conversational dialogue system. To create a coherent and engaging dialogue system, the context-aware topic classification method with dialogue act helps in unsupervised topic keyword detection [7]. Topic tracking in conversational dialogue [17], semantic similarity, making an evaluation, continuous state tracking, Multi-functional behaviour and more unsupervised learning [14] is also big challenges in dialogue systems.

## 3. RELATED WORK

The original idea of topic detection originated in 1996 as part of its broadcast news at the US Government Defence Advanced Research Projects Agency (DARPA) [2]. Detecting topics can be valuable as soon as possible to discover natural disasters [4, 9], helping political parties to predict election results [12] and companies to understand user opinions to improve marketing contents for better understanding of customers' needs [11]. One of the common representations is describing each topic by a set of keywords. This set can be a weighted set of keywords, where weights represent the keywords and their importance in the topic.

Topic detection from textual data falls in three categories: document-pivot, feature-pivot approaches and probabilistic methods [1]. Firstly, document pivot method groups the individual documents according to the document similarity. Secondly, feature pivot method groups together the terms according to their co-occurrence features pattern. Lastly, probabilistic models treat the problem of topic detection as a probabilistic inference problem. Many techniques have been proposed for topic detection, including clustering, frequent pattern mining, matrix factorisation and exemplar-based topic detection. These existing approaches are used to detect topics from tweets [21], textual documents such as Wikipedia [15] and textual blogging [13]. Clustering involves the organizing of objects into meaningful groups known as a cluster.Objects in one cluster would likely be different from objects grouped under another cluster. The centroid is used as a representative for each cluster discovered, where the top t words (in terms of TF-IDF weights) are used as the keywords of this topic. To detect topics, each utterance in the dialogue is represented using TF-IDF scheme and the number of topics to be discovered is used as the number of clusters (k). The combination of k-means clustering and elbow method improves the efficiency and effectiveness of k-means performance in processing a large amount of data [6]. Incremental clustering with vector expansion extracts scores automatically and utilizes temporarily term similarities for online event detection in microblogs [18]. They use temporal context in microblog posts to detect similar terms by using incremental clustering techniques. When corpus contains closely related topics, then feature pivot approach to detect co-occurrence patterns simultaneously for a large number of terms perform better for textual topic detection [19].

In the human-machine conversation, an engaging and coherent response is possible if the context of the conversation is taking into account. Deep average network and attention deep average network explore various features to incorporate the context of the conversation and dialogue acts gain in topic classification accuracy by 35% and unsupervised keyword detection recall function by 11% [20]. The Latent Dirichlet Allocation (LDA) [3] idea was a mix of topics in which each topic performs as a latent multinomial variable characterized by a distribution of words over a defined vocabulary.

## 4. PROPOSED METHOD

In the previous work, there are multiple approaches for topic detection such as topic detection with clustering techniques, frequent pattern mining, exemplar-based approach, matrix factorisation and probabilistic models. In our approach, firstly, we combined term similarity analysis by analysing frequent pattern in the dataset to detect topics and k-means clustering to make clusters for all high-frequency words in topics. Secondly, LDA topic model combined with elbow method to select the optimal number of clusters. In the experiment, topic detection is divided into three sections. Data pre-processing, term similarity analysis with clustering and elbow method and topic detection with Parallel Latent Dirichlet Analysis (PLDA). Figure 1 shows the proposed method.

### 4.1. Data Pre-processing

For the experimental purpose, we use switchboard corpus. Switchboard is a set of approximately 2,400 two-sided telephone conversations between 543 speakers (302 male, 241 female) from all over the United States. We used total 2145 conversation and removed smaller conversations such as "uh-huh", "okay", "right", "oh", "um-hum" etc. In the experiment, we used unsupervised data, only the dialogue utterances.

- Markup Tag Filter: From the input column, removed all markup language tags.

- Stanford Tagger: assigns to each term of a document a part of speech (POS) tag.

- Punctuation Eraser: From the input documents, removed all the punctuation characters.

- Number filter: From the input documents, filtered all terms contains digits, including decimal separators "," or "." and possible leading "+" or "-".

- N Char Filter: Filters all terms contained in the input documents with less than the specified number N characters (we set the value 3).

- Stop Word Filter: Removed stop words and filtered all the input documents.

- Case Convertor: In the input documents, convert all the term to lower or upper case.

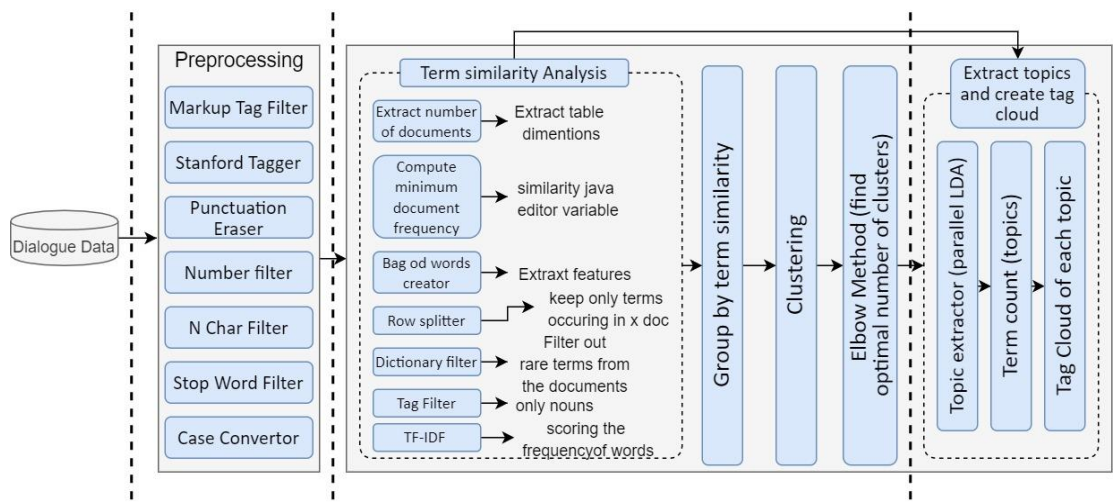

Figure 1. Proposed Method for Topic Detection

### 4.2. Term Similarity Analysis

The first challenge for topic detection within a conversational dialogue is to find the dialogue utterances that are similar in content, under the term similarity analysis. The dialogue is composed of utterances, and each utterance is considered as a single document. For each row, a document will be created and attached to that row to extract the number of rows for the table dimensions. The Bag of word (BoW) model is used to extract the features from each document. It collects the data in strings and designs the vocabulary of known words. TF-IDF scores the frequency of the words in the current document. It is also scoring of how rare the words are across documents. The concept of BoW and TF-IDF is necessary to train the PLDA model. Dictionary filter filters the high-frequency words from the documents. By applying simple k-means, an unsupervised machine learning algorithm that groups all the high-frequency words into k number of clusters. The elbow method helps the interpretation and validation of consistency within-cluster analysis and select the optimal number of clusters by fitting the model with a range of values for K.

Computer Science & Information Technology (CS & IT) 99

### 4.3. Topic Extraction and making Tag Clouds

The LDA model defined as a generative probabilistic model for collections of discrete data such as text corpora. It imagines a fixed set of topics. Each topic represents a set of words. LDA's goal is to map all documents to relevant topics in such a way that those imaginary topics mostly capture the words in each document. The concept behind LDA is that each document can be represented through a topic distribution and each topic can be described through a word distribution, which is the premise of the ' bag of words '. In our approach LDA is taking input parallel from elbow method and term similarity based on BoW and TF-IDF to compute the topics.

## 5. EXPERIMENTAL RESULTS

To determine the optimal clusters in the corpus is the fundamental issue in the clustering technique. The elbow method looks the total within clusters sum of square (WSS) error and minimizes this to absolute value. The optimal number of clusters is defined as follows in the Figure 2:

1. To calculate the clustering algorithm (e.g., clustering of k-means) for different k values varies from 1 to 10. It also depends on the term similarity of word frequency value.
2. Calculate the total within-cluster sum of square (WSS) error value for each k.
3. Plot the elbow method WSS curve based on the number of clusters k.
4. The curve in the graph is generally considered an indication of the optimal number of clusters.

From the experimental results, there are a total of 20 optimal clusters, and each cluster contains a different number of similar items. The elbow method selects the optimal number of clusters in Figure 2 and decreases the sum of squared errors within the clusters in Figure 2.

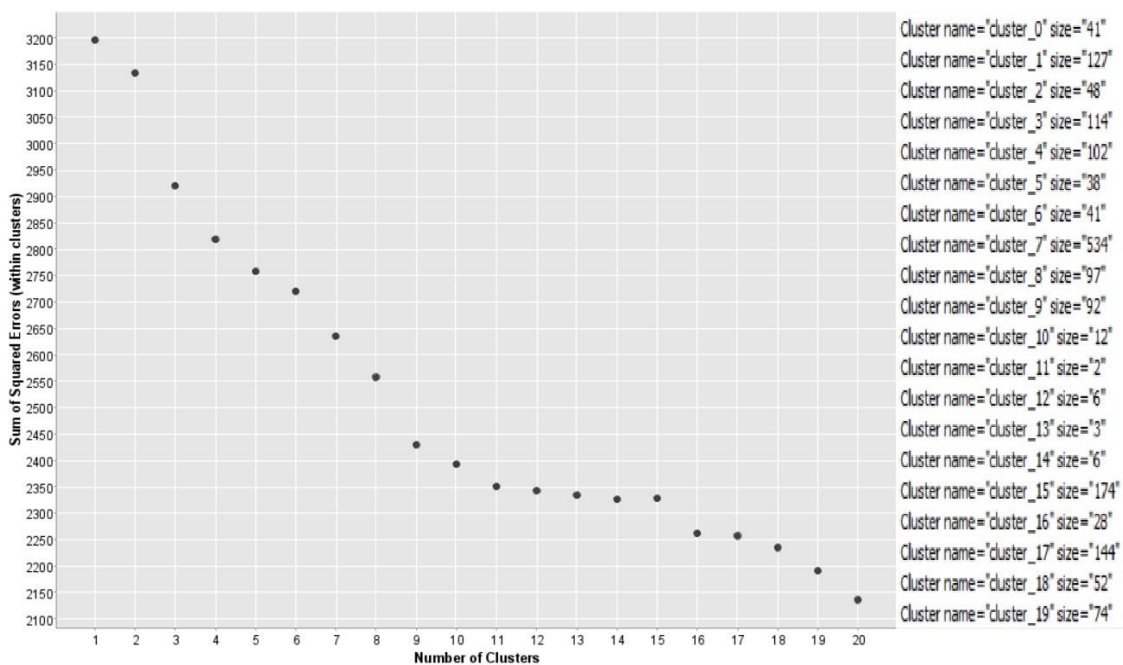

Figure 2. Optimal numbers of clusters from Elbow Method and WSS curve

The PLDA model detects the topics with parameters alpha is 0.5, beta 0.1 and sampling iteration 1000. The number of topics is 3, with ten words in each topic and representation of tag clouds in Figure 3.

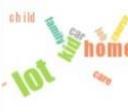

Figure 3. Representations of topics in tag clouds

We reformulated the problem in terms of standard information retrieval evaluation metrics:
*Precision = PP/PNr, Recall = PP/NP, and*
*F-measures = [2(precision)(recall)]/ [(precision + recall)(PP)]*

Table 1. Performance comparison between different methods with alpha 0.5 and beta 0.1

| Methods | Precision | Recall | F-measures |
|---|---|---|---|
| LDA | 0.762 | 0.834 | 0.874 |
| Clustering (K-means) | 0.778 | 0.861 | 0.899 |
| PLDA+Elbow method | 0.846 | 0.931 | 0.915 |

## 6. CONCLUSIONS

On this work we proposed a topic detection approach combining with clustering, elbow method and PLDA model. The first step in a series of different approached to refine the dataset. Then the data is being used to extract similar features word and transform in to multiple clusters. The elbow method with clustering interprets and validates the consistency within the clusters and select optimal number of clusters. LDA is taking input parallel from elbow method and term similarity based on BoW and TF-IDF to compute the topics. We compare our approach with simple LDA model and clustering to evaluate precision, recall and F-measures.

**ACKNOWLEDGEMENTS**

This research was conducted with the financial support of the Science Foundation Ireland under grant agreement No. (13/RC/2106) at the ADAPT SFI Research Centre at Trinity College Dublin, University of Dublin, Ireland. The ADAPT SGI Centre for Digital Media Technology is funded by Science Foundation Ireland through the SFI Research Centre Programmes and is Co-funded under the European Regional Development Fund (ERDF) through grant #13/RC/2016.

Computer Science & Information Technology (CS & IT)                          101

**AUTHORS**

Haider Khalid is an ADAPT PhD student at Trinity College Dublin, Ireland. His current research is focused on topic detection and topic modelling for dialog system with NLP. He holds a Master's degree in Software Engineering from Jiangsu University China, funded by Chinese Scholarship Council (CSC). Outside of the office, he likes sport, socializing with people and nature. He is also very interested in politics, technology and different cultures.

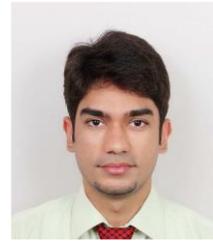

Professor Vincent Wade is Director of the ADAPT Centre for Digital Media Technology and holds the Professorial Chair of Computer Science (Est. 1990) in School of Computer Science and Statistics, Trinity College Dublin as well as a personal Chair in Artificial Intelligence. His research focuses on intelligent systems, AI and Personalisation. He was awarded Fellowship of Trinity College for his contribution to research and has published over three hundred and fifty scientific papers in peer reviewed international journals and conferences.